\newif\ifpreprint
\preprintfalse % Enable for double column reprint

\ifpreprint
\documentclass[journal=jpclcd,manuscript=letter]{achemso}
\else
\documentclass[journal=jpclcd,manuscript=letter,layout=twocolumn]{achemso}
\fi

\usepackage[T1]{fontenc} % Use modern font encodings

\usepackage{amsmath}
\usepackage{newtxtext,newtxmath}

\usepackage{pifont}
\usepackage{graphicx}
\usepackage{dcolumn}
\usepackage{braket}
\usepackage{multirow}
\usepackage{threeparttable}
\usepackage{xspace}
\usepackage{verbatim}
\usepackage[version=4]{mhchem} % Formula subscripts using \ce{}
\usepackage{comment}
\usepackage{color,soul}
\usepackage{booktabs} 
\usepackage{mathtools}
\usepackage[dvipsnames]{xcolor}
\usepackage{xspace}
\usepackage{ifthen}

\usepackage{qcircuit}

\usepackage{graphicx,longtable,dcolumn,mhchem}
\usepackage{rotating,color}
\usepackage{lscape}
\usepackage{amsmath}
\usepackage{dsfont}
\usepackage{soul}
\usepackage{physics}

\usepackage[colorlinks = true,
            linkcolor = blue,
            urlcolor  = black,
            citecolor = blue,
            anchorcolor = black]{hyperref}
\definecolor{goodorange}{RGB}{225,125,0}
\definecolor{goodgreen}{RGB}{5,130,5}
\definecolor{goodred}{RGB}{220,50,25}
\definecolor{goodblue}{RGB}{30,144,255}

\newcommand{\note}[2]{
\ifthenelse{\equal{#1}{F}}{
\colorbox{goodorange}{\textcolor{white}{\footnotesize \fontfamily{phv}\selectfont #1}}
    \textcolor{goodorange}{{\footnotesize \fontfamily{phv}\selectfont #2}}\xspace
}{}
\ifthenelse{\equal{#1}{R}}{
\colorbox{goodred}{\textcolor{white}{\footnotesize \fontfamily{phv}\selectfont #1}}
    \textcolor{goodred}{{\footnotesize \fontfamily{phv}\selectfont #2}}\xspace
}{}
\ifthenelse{\equal{#1}{N}}{
\colorbox{goodgreen}{\textcolor{white}{\footnotesize \fontfamily{phv}\selectfont #1}}
    \textcolor{goodgreen}{{\footnotesize \fontfamily{phv}\selectfont #2}}\xspace
}{}
\ifthenelse{\equal{#1}{M}}{
\colorbox{goodblue}{\textcolor{white}{\footnotesize \fontfamily{phv}\selectfont #1}}
    \textcolor{goodblue}{{\footnotesize \fontfamily{phv}\selectfont #2}}\xspace
}{}
}

\usepackage{titlesec}

\usepackage[fontsize=11pt]{scrextend}
\titleformat{\section}
{\normalfont\sffamily\bfseries\color{Blue}}
{\thesection.}{0.25em}{\uppercase}

\titleformat{\subsection}[runin]
{\normalfont\sffamily\bfseries}
{\thesubsection}{0.25em}{}[.\;\;]

\titleformat{\suppinfo}
{\normalfont\sffamily\bfseries}
{\thesubsection}{0.25em}{}

\titlespacing*{\section}{0pt}{0.5\baselineskip}{0.01\baselineskip}
\titlespacing*{\subsection}{0pt}{0.125\baselineskip}{0.01\baselineskip}

\author{Miroslav Kolos}
	\affiliation[NEEL, Grenoble]{Department of Physics, Faculty of Science, University of Ostrava, 30.~dubna 22, 701 03 Ostrava, Czech Republic}
\author{František Karlický}
    \email{frantisek.karlicky@osu.cz}
	\affiliation[CEA, Grenoble]{Department of Physics, Faculty of Science, University of Ostrava, 30.~dubna 22, 701 03 Ostrava, Czech Republic}

\let\oldmaketitle\maketitle
\let\maketitle\relax
	\title{Predicting Fundamental Gaps of Chromium-Based 2D Materials Using GW Methods}
\date{\today}

\begin{tocentry}
\includegraphics[width=0.99\columnwidth]{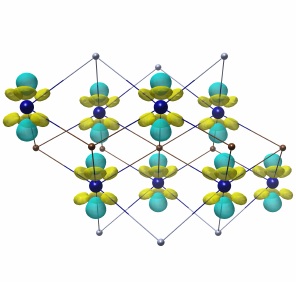}
\end{tocentry}
\RequirePackage[normalem]{ulem} %DIF PREAMBLE
\RequirePackage{color}\definecolor{RED}{rgb}{1,0,0}\definecolor{BLUE}{rgb}{0,0,1} %DIF PREAMBLE
\RequirePackage{listings} %DIF PREAMBLE
\RequirePackage{color} %DIF PREAMBLE
\lstdefinelanguage{DIFcode}{ %DIF PREAMBLE
  moredelim=[il][\color{red}\sout]{\%DIF\ <\ }, %DIF PREAMBLE
  moredelim=[il][\color{blue}\uwave]{\%DIF\ >\ } %DIF PREAMBLE
} %DIF PREAMBLE
\lstdefinestyle{DIFverbatimstyle}{ %DIF PREAMBLE
	language=DIFcode, %DIF PREAMBLE
	basicstyle=\ttfamily, %DIF PREAMBLE
	columns=fullflexible, %DIF PREAMBLE
	keepspaces=true %DIF PREAMBLE
} %DIF PREAMBLE
\lstnewenvironment{DIFverbatim}{\lstset{style=DIFverbatimstyle}}{} %DIF PREAMBLE
\lstnewenvironment{DIFverbatim*}{\lstset{style=DIFverbatimstyle,showspaces=true}}{} %DIF PREAMBLE
\begin{document}	

\ifpreprint
\else
\twocolumn[
\begin{@twocolumnfalse}
\fi
\oldmaketitle

%%%%%%%%%%%%%%%%
%%% ABSTRACT %%%
%%%%%%%%%%%%%%%%

\begin{abstract}
Precise and accurate predictions of the two-dimensional (2D) material's fundamental gap are crucial for next-generation flexible electronic and photonic devices. We, therefore, evaluated the predictivity of the GW approach in its several variants built on various density functional theory (DFT) inputs. We identified reasons for significant discrepancies between generalized gradient approximation and hybrid DFT results for intricate cases of 2D materials containing chromium and evaluated the diverse behavior of subsequent quasiparticle corrections. We examined the impact of omitted vertex corrections using the more computationally intensive quasiparticle self-consistent QPGW and partially self-consistent QPGW$_0$ methodologies. We observed consistent trends across Cr-based and other 2D materials compared by advanced GW calculations, suggesting that single-shot G$_0$W$_0$@PBE can provide reasonable estimates of fundamental gaps when applied with caution. While this approach shows promise for a variety of 2D materials, including complicated antiferromagnetic chromium-based transition metal carbides (MXenes), further research is required to validate its reliability for strongly correlated systems. In contrast, the G$_0$W$_0$@HSE06 approach may strongly overestimate gaps in some cases.
\end{abstract}

\ifpreprint
\else
\end{@twocolumnfalse}
]
\fi

\ifpreprint
\else
\small
\fi

\noindent

%%%%%%%%%%%%%%%%%%%%
%%% INTRODUCTION %%%
%%%%%%%%%%%%%%%%%%%%
\section{Introduction}
The fundamental gap is an important property of any new material. 
Namely, gaps in semiconducting and insulating materials have been of critical importance for applications and industry for a long time,\cite{Capasso1987} and reliable predictions of fundamental gaps are therefore needed. 
Density functional theory (DFT) approximates the fundamental gap (defined using ionization energy and electron affinity) by the electronic gap from Kohn-Sham eigenvalues.\cite{Perdew2017} 
As local and semilocal DFT strongly underestimate the band gaps of semiconductors and insulators due to self-interaction error,\cite{MoriSanchez2008} the band gap problem is solved by many-body methods or often effectively overcome by hybrid density functionals.\cite{Perdew2017} 
The same is valid in recent research on a new generation of atomically thin films of semiconducting and insulating materials, two-dimensional (2D) materials, with the unique potential to fabricate flexible and ultrathin devices.\cite{Chaves2020} 
Despite that DFT and many-body methods behave quite standardly in 2D materials, in our recent investigations\cite{Ketolainen2022}, we discovered a chromium puzzle when we explored the electronic properties of transition metal carbides (MXenes), specifically Cr$_2$CF$_2$ and Cr$_2$C(OH)$_2$.
First of all, the hybrid HSE06 density functional\cite{Heyd2003} surprisingly provided an electronic band gap value of more than 300\% of the one from generalized gradient approximation (GGA) of DFT by Perdew-Burke-Erzenhof (PBE).\cite{Perdew1996} 
Such a significant difference is not typically observed\cite{Civalleri2012,Garza2016}. 
This discrepancy highlights potential limitations in conventional computational methods for studying transition-metal-based systems like those considered here. 
Further, our careful studies (see below) on complicated semiconducting antiferromagnetic (AFM) Cr$_2$CF$_2$ MXene using many-body GW approximation\cite{Hedin1965} in time-proven single-shot G$_0$W$_0$@PBE version revealed a band gap that is unexpectedly lower than that from hybrid HSE06 calculations. % (for structure, stability, and details on Cr$_2$CF$_2$, see Supporting Information (SI)). 
Given that GW methods are approximations of Hedin's exact equations\cite{Hedin1965} and depend on the input wavefunction quality, we conducted further studies to investigate these disparities between PBE and HSE06 density functionals to clarify the unusual behavior observed in MXenes and check general properties of hybrid functionals and GW methods for 2D materials. Converged fundamental gaps from G$_0$W$_0$@PBE are considered as a reference, and band gaps obtained from HSE06 on the other hand are considered as a practical compromise. We aim to test whether these considerations are valid for such complicated materials. 

\section{Computational Details}
%\textit{Methods:} 
Our study utilizes the Vienna ab initio simulation package (VASP) with the projector augmented wave (PAW) method for comprehensive electronic structure calculations\cite{Kresse1999, Blochl1994}. We employ GW-specific PAWs, treating $s$ and $p$ electrons explicitly, with $d$ electrons included for chromium and scandium.
Atomic positions are optimized using density functional theory (DFT) with the Perdew-Burke-Ernzerhof (PBE) functional\cite{Perdew1996}, applying the conjugate gradient algorithm until forces are below $10^{-3}$ eV/Å. Electronic steps achieve self-consistent field (SCF) convergence at $10^{-7}$ eV, with a plane-wave basis set energy cut-off of 400 eV. Lattice vibrational frequencies for Cr$_2$CF$_2$ are calculated using PBE within Phonopy\cite{phonopy}, with a 0.01 Å displacement and an 8×8×1 supercell (section S1 in SI).
To enhance electronic structure accuracy, we incorporate the HSE06 hybrid functional\cite{Heyd2003}, which combines Hartree-Fock exact exchange with PBE exchange-correlation. We used the G$_0$W$_0$ approximation, involving Green's function $G$, and screened Coulomb potential $W$ with input orbitals from PBE density functional (see section S2 in SI for settings and convergences) and HSE06 density functional.  
Our approach also includes full self-consistent quasiparticle QPGW and QPGW$_0$ calculations, updating wavefunctions at each iteration.\cite{Grumet2018} $G$ in QPGW$_0$ and both $G$ and $W$ in QPGW are recalculated using updated quasiparticle energies and refined wavefunctions (section S3 in SI).  
Quasiparticle gaps were subtracted from conduction band minimum (CBM) and valence band maximum (VBM) energies.

\section{Results and discussion}
Before investigating the discrepancies between PBE and HSE06 observed in Cr$_2$CF$_2$, we first studied its stability and preferred structural conformer. Our analysis found that the preferred and stable structure is trigonal (particularly trigonal type 2; for details, see the Supporting Information (SI)). To gain a deeper understanding of the discrepancies between PBE and HSE06, we expanded our investigation and based our considerations on three 2D materials: two simpler ($s$,$p$-containing elements), hexagonal boron nitride (BN) and fluorographene (CF), and one more difficult scandium-based MXene Sc$_2$C(OH)$_2$. 
These 2D materials were selected because their carefully converged electronic G$_0$W$_0$@PBE band gaps are in perfect agreement with experiment (layered bulk BN)\cite{Kolos2019} and quantum Monte Carlo (QMC) (CF and  Sc$_2$C(OH)$_2$).\cite{Dubecky2020, Dubecky2023} 
These materials' precise and accurate G$_0$W$_0$@PBE fundamental gaps are presented in Table \ref{tab:bandgap}. 
We performed the same procedure of careful convergence of technical parameters in G$_0$W$_0$@PBE for Cr$_2$CF$_2$ (see SI for details), and after considerations introduced later in this paper, we added band gap to this set too.
This selection of 2D materials aims to confirm the predictive accuracy of G$_0$W$_0$@PBE for complicated chromium-based MXenes, enhancing our understanding of the electronic properties of diverse 2D materials.

\begin{table}[ht]
 \begin{threeparttable}
\footnotesize
\caption{Precise G$_0$W$_0$@PBE direct (indirect) band gap values of four 2D materials in eV.\label{tab:acc_GW}}
\label{tab:bandgap}
\setlength\tabcolsep{2.0pt}
\begin{tabular}{@{}llll@{}}
\hline
\toprule
Material & G$_0$W$_0$ Gap & Experiment/QMC & Reference \\ \midrule
%bulk BN  & 6.53 & Kolos \textit{et al.}\cite{Kolos2019} \\
BN$^\textbf{a}$ & 6.53 (6.08) & 6.42\cite{Doan2016} (6.08 $\pm$ 0.015\cite{Cassabois2016}) & Kolos \textit{et al.} \cite{Kolos2019} \\
CF  & 7.14& 7.1 $\pm$ 0.1 & Dubecký \textit{et al.}\cite{Dubecky2020}\\
Sc$_2$C(OH)$_2$ & 1.28 & 1.30 $\pm$ 0.2  & Dubecký \textit{et al.}\cite{Dubecky2023}  \\
Cr$_2$CF$_2$  &  2.41 (2.29) & $\times$ &This work \\ \bottomrule
\hline
\end{tabular}
\begin{tablenotes}
     \scriptsize
\item $^\textbf{a}$ $\mathrm{AA^\prime}$ stacking of layered BN; precise G$_0$W$_0$@PBE band gap for monolayer BN  is 7.64 (6.94) from Kolos \textit{et al.}\cite{Kolos2022}.
\end{tablenotes}
\end{threeparttable}
\end{table} 
Our primary focus starts with understanding the essence of the vast differences between band gaps obtained from GGA PBE and hybrid HSE06 density functionals in chromium MXenes (e.g., Cr$_2$CF$_2$ indirect gap of 1.11 eV vs. 3.27 eV, respectively) because subsequent quasiparticle corrections are dependent on input DFT wave functions and eigenvalues.
The discrepancies between one-electron methods can be associated with chromium's half-filled 3d orbitals, which exhibit a complex electronic landscape with complicated electron-electron interactions. Standard DFT methods use exchange-correlation functionals such as PBE to describe complicated electron-electron interactions. PBE, thanks to a good approximation of electronic screening, is believed to give reasonable values of macroscopic dielectric constant\cite{Petousis2016}. On the other hand, HSE06 is a range-separated hybrid functional combining a short-range component of 25\% Hartree-Fock exact exchange with 75\% PBE exchange while retaining the full PBE correlation. The differences in macroscopic dielectric constant from PBE and HSE06 are vast in chromium MXene, up to nearly three times, compared to other studied 2D materials, where there are just minor changes (Tab. \ref{tab:eps}). This first hint urges a deeper study of PBE and HSE06 results for these chromium-based 2D materials. 

\begin{table}[]
\footnotesize
 \begin{threeparttable}
\caption{Macroscopic dielectric constants for various 2D materials in various theory levels.$^\textbf{a}$}
\label{tab:eps}
\setlength\tabcolsep{2.0pt}
\begin{tabular}{lccccc}
\hline
\toprule
         & BN   & CF   & Sc$_2$C(OH)$_2$ & Cr$_2$CF$_2^\mathrm{AFM}$ & Cr$_2$CF$_2^\mathrm{NM}$ \\ \midrule
PBE      & 1.51 & 1.68 & 4.85   & 7.68 & 16.24 \\
HSE06     & 1.38 & 1.56 & 3.14   & 3.18 & 5.69 \\
QPGW$_0$@PBE & 1.52 & 1.69 & 4.49   & 5.58 & 18.61 \\
QPGW$_0$@HSE06& 1.40 & 1.56 & 3.02   & 2.84 &  6.11  \\
QPGW       & 1.33 & 1.46 & 2.68   & 2.83 &  8.70\\\bottomrule
\hline

\end{tabular}
\begin{tablenotes}
     \scriptsize
\item $^\textbf{a}$ All presented 2D materials are calculated using cell size of $\approx$ 20\r{A} in the z-direction because, in the $\infty$ limit, these values are always 1.\cite{Ketolainen2020} This table, therefore, serves to compare different approaches.
\end{tablenotes}
\end{threeparttable}
\end{table}

We looked at charge density itself to understand what caused such differences between two density functionals. 
\begin{figure}[bht]
\includegraphics[width=0.99\columnwidth]{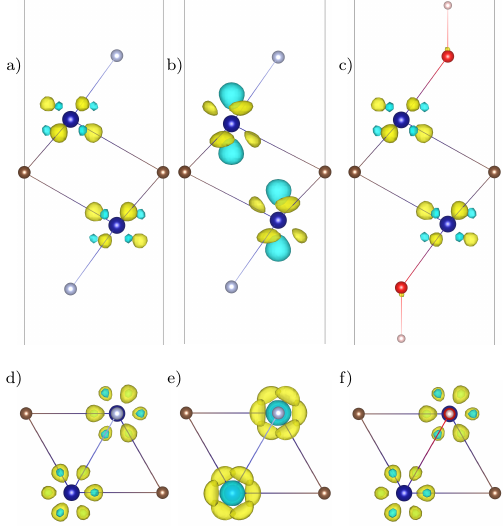}
\caption{\label{fig:CHG} Differences in charge densities from PBE and HSE06 density functionals ($n_\mathrm{PBE}-n_\mathrm{HSE06}$) at the same isosurface level ($3.5\mathrm{\cdot10^{-5}}e$) as side and top views of a),d) Cr$_2$CF$_2^\mathrm{AFM}$, b),e) Cr$_2$CF$_2^\mathrm{NM}$, and d),f) Cr$_2$C(OH)$_2^\mathrm{AFM}$, all with highlighted unit cell boundaries in a)-c). Atom colors: brown-C, blue-Cr, grey-F, red-O, white-H. All other studied 2D materials have no visible differences at the same isosurface level. }
\end{figure}
\begin{figure*}[h]
\includegraphics[width=1.99\columnwidth]{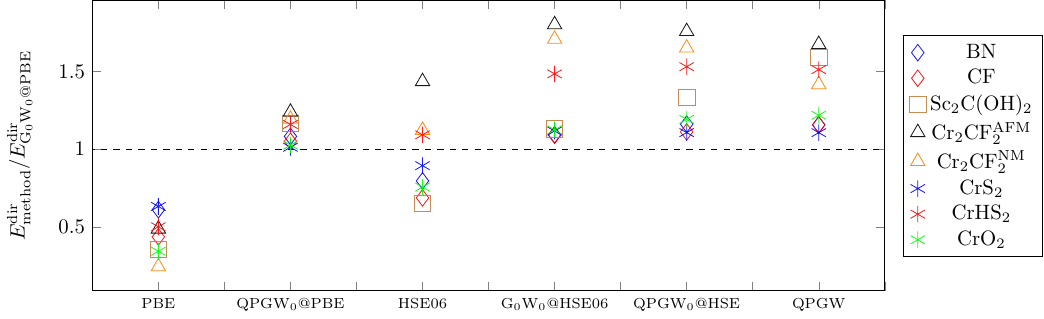}
\caption{\label{fig:dots}  Fundamental gaps of several 2D materials obtained from different methods (abbreviated in x-axis labels) divided by the corresponding G$_0$W$_0$@PBE fundamental gap, presented as a horizontal dashed line. }
\end{figure*}
A striking observation was the substantial differences in electron density between the PBE and HSE06 functionals, as depicted in Figure \ref{fig:CHG}. 
No such differences were observed in other studied 2D materials in our set (BN, CF, and Sc$_2$C(OH)$_2$) on the same isosurface level. To determine whether these discrepancies could be attributed to variations in magnetic moments -- a known difference between PBE and HSE06\cite{Meng2016} -- we examined both magnetic and nonmagnetic (NM) (open-shell and closed-shell) forms of Cr$_2$CF$_2$. Remarkably, even in nonmagnetic configurations, notable differences in charge densities persisted (Fig. \ref{fig:CHG}), indicating that these variations are intrinsic to the chromium atoms rather than influenced by magnetic properties. This is also supported by significant differences in Hartree energy between PBE and HSE06 (Tab. S2).

In order to assess whether the termination groups could influence these discrepancies, we analyzed Cr$_2$C(OH)$_2$, another antiferromagnetic material (Fig. \ref{fig:CHG}). The results clearly demonstrated that the observed differences in electron density are explicitly associated with the chromium atoms, confirming that the termination groups do not contribute to the observed discrepancies. Expanding the scope of our investigation into the charge density HSE06/PBE differences, we included MXenes containing titanium and manganese elements, each exhibiting distinct magnetic properties. In all cases, no differences in charge density between PBE and HSE06 were observed at the same isosurface levels, reinforcing chromium's unique challenge in accurately modeling electron density. We did an integration of its absolute values to quantify the differences in charge densities. We used a threshold value for differences included in the integration, consistent for all materials to include just these high-order differences and exclude vacuum noise. These results are presented in Table S3, and it is visible that MXenes containing chromium have about one order of magnitude higher differences than other materials. Similar behavior of charge density differences between PBE and HSE06, but less pronounced, was also observed in simpler 2D materials containing chromium from families of transition metal dichalcogenides and oxides (CrS$_2$, CrHS$_2$, and CrO$_2$; Fig. S3). It is visible that these less correlated chromium 2D materials show more minor absolute differences in Bader charge analysis (Tab. S3).
Such an unexpected behavior of charge density suggests evaluating electron-electron interactions in chromium MXenes on a higher level of theory.

To address the inconsistencies in electron-electron interaction modeling in chromium-based 2D materials compared to other 2D materials, we adopted the many-body GW method, which more accurately accounts for these interactions, yet without including vertex corrections\cite{Grumet2018}. Specifically, we explored various levels of GW, including the standard single-shot corrections G$_0$W$_0$, and the more sophisticated quasiparticle self-consistent QPGW and partially self-consistent QPGW$_0$ methods. Notably, all these iterative QPGW approximations omit vertex corrections as specified in Hedin's equations, which leads to the systematic overestimation of band gaps due to the absence of part of the electron-electron interactions in the self-consistent versions\cite{Grumet2018}.

Our findings are summarized in Figure \ref{fig:dots}, showing the relative differences in calculated fundamental gaps from the baseline G$_0$W$_0$@PBE (values from Tab. \ref{tab:bandgap}), with further methodological discussion available in the Supplementary Information. One striking result is the significant discrepancy between the HSE06-based results for the Cr$_2$CF$_2$ band gap and all other outcomes, highlighting the limitations of using HSE06-derived input wave functions for GW calculations. This indicates a divergence in the treatment of electron-electron interactions in chromium-based MXenes.

Importantly, our findings reveal that the behavior of relative differences between QPGW$_0$@PBE and G$_0$W$_0$@PBE closely mirrors across all tested 2D materials, including the problematic Cr$_2$CF$_2$. QPGW$_0$ methods use screened Coulomb potential W$_0$ calculated from non-interacting Green's function on input orbitals (PBE, HSE06). Full self-consistency is achieved by updating energies and wavefunctions based on Green's repeated propagation through the system with PBE/HSE06-based screened coulomb interaction. This method, therefore, shows how reasonable the input orbitals were, and therefore, PBE input orbitals seem to be reasonable regardless of the system. Noting that G$_0$W$_0$@PBE band gap results for BN, CF, and Sc$_2$C(OH)$_2$ were proven to be precise and accurate, and aforementioned similar behavior of relative differences between G$_0$W$_0$@PBE and GW$_0$@PBE for all studied materials (Cr$_2$CF$_2$ including), this could be interpreted that G$_0$W$_0$@PBE provides reliable predictions currently achievable without vertex corrections, even for complicated chromium-based 2D materials. However, due to the lack of experimental or high-level quantum chemistry calculations specifically supporting the precision of G$_0$W$_0$@PBE for Cr$_2$CF$_2$, caution still should be exercised when generalizing its reliability for strongly correlated systems. Further research, including experimental validations and advanced computational studies, is necessary to confirm these findings for chromium-based 2D materials and other complex systems.

A crucial aspect of our research involves thoroughly examining the QPGW results. 
As illustrated in Figure \ref{fig:dots}, the fully self-consistent QPGW method tends to significantly overestimate the band gaps of strongly correlated or magnetic $d$-element-based 2D materials studied, underscoring the pivotal role of complicated electron-electron interactions. These effects are not accounted for due to the absence of vertex corrections in the GW framework, which appears particularly critical in such types of 2D materials, intensifying the overestimations. On the other hand, less correlated NM Cr materials, CrS$_2$ and CrO$_2$, closely mirror $sp$-element-based 2D materials.

\begin{figure}[ht]
\includegraphics[width=0.79\columnwidth]{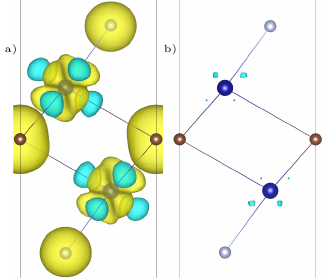}
\caption{\label{fig:CHG2} Differences in charge densities in Cr$_2$CF$_2$ a) $n_\mathrm{QPGW_0@PBE}- n_\mathrm{QPGW@PBE}$ and b) $n_\mathrm{QPGW_0@HSE06} - n_\mathrm{QPGW@HSE06}$, at the same isosurface level ($3.5\mathrm{\cdot10^{-6}}e$).}
\end{figure}

Our focus switches to analyzing band gap results obtained from G$_0$W$_0$@HSE06 and QPGW$_0$@HSE06. The results from these calculations, including QPGW, demonstrated surprisingly similar band gaps for both NM and AFM Cr$_2$CF$_2$, with the self-consistent approaches yielding even smaller band gaps than one observed in the single-shot G$_0$W$_0$. In principle, QPGW results overestimate band gaps;\cite{Grumet2018} therefore, from our results, it is clear that HSE06 orbitals are unsuitable inputs for subsequent many-body G$_0$W$_0$ calculations for these materials because G$_0$W$_0$@HSE06 gaps are even larger than QPGW ones. This phenomenon is further explored by comparing charge densities from GW and GW$_0$, using PBE and HSE06 orbitals, as detailed in Figure \ref{fig:CHG2}. Surprisingly, iterations of W starting from W$_0$ obtained from HSE06 orbitals show much smaller differences in charge density than those from PBE. 
This investigation led us to conclude that W$_0$, derived as a single-shot screened Coulomb interaction from a non-interacting Green's function propagated through the HSE06 wave function, mimics a fully converged W. This behavior, on the one hand, proposes the credibility of HSE06, on the other hand, considering shortcoming of QPGW leading to fundamental gap overestimations (because of omitting vertex corrections),\cite{Grumet2018} serves as evidence of HSE06's inadequacy as input orbitals for GW calculations for predicting the electronic structure and band gaps of such highly correlated chromium 2D materials. 
In addition, HSE06 overestimates magnetic moments\cite{Ketolainen2022}, can fail because of multireference system\cite{Cho2023} or due to tradeoffs between over-delocalization and under-binding,\cite{Janesko2021} and subsequent quasiparticle G$_0$W$_0$@HSE06 gaps often overestimate experiments in strongly correlated materials, as shown e.g. for antiferromagnetic hematite $\alpha$-\ce{Fe2O3}.\cite{Liao2011} 
It is important to note, however, that the G$_0$W$_0$@HSE06 approach is particularly valuable in the case of small-gap semiconductors, where PBE provides (artificially) negative gap and subsequent perturbative G$_0$W$_0$@PBE is not available.\cite{Grumet2018} 
Conversely, our findings highlight the remarkable efficiency of the PBE density functional, which consistently provides highly reliable results when used as the input for single-shot GW calculations in these challenging 2D compounds.

From our results in Figure \ref{fig:dots}, it is evident that the relative variability in GW-calculated band gaps significantly depends on the material's electronic and magnetic properties. For nonmagnetic, weakly correlated materials such as BN, CF, CrS$_2$, and CrO$_2$, the maximal differences in band gaps across different levels of GW (e.g., QPGW$_0$, QPGW) and input orbitals do not exceed 20\%. However, when considering systems with magnetic properties (e.g., CrHS$_2$) or stronger correlations (e.g., Sc$_2$C(OH)$_2$, Cr$_2$CF$_2^{\mathrm{NM}}$), or the combination of both (Cr$_2$CF$_2^{\mathrm{AFM}}$), these discrepancies increase significantly, ranging from 50 \% to 80 \% of the band gap. These findings highlight the need for careful consideration of GW methodologies, particularly for magnetic and strongly correlated materials, where variations in band gap predictions can be substantial.

The evaluation of the macroscopic dielectric constant across various 2D materials further underscores the reliability of the G$_0$W$_0$@PBE method even for Cr$_2$CF$_2$. As presented in Table \ref{tab:eps}, there is a substantial discrepancy between the dielectric constants calculated using PBE and HSE06 for Cr$_2$CF$_2$, with PBE predicting a notably higher value compared to HSE06. This significant variance suggests that the choice of exchange-correlation functional profoundly impacts the predicted dielectric properties, especially in 2D materials with strong electronic interactions such as chromium-based MXenes. Interestingly, the values obtained from GW$_0$@PBE are between those of PBE and HSE06, whereas GW$_0$@HSE06 closely aligns with the HSE06 values; note that results of the dielectric constant from G$_0$W$_0$ and GW$_0$ are the same, due to the same response function. This behavior provides additional validation for using G$_0$W$_0$@PBE as a robust method, capturing crucial aspects of material behavior.

\section{Conclusions}
The most significant finding from our data is that the fundamental gap of chromium MXene Cr$_2$CF$_2$ is wider in both its magnetic and nonmagnetic forms when calculated with G$_0$W$_0$@HSE06 compared to the results from QPGW. Highlighting a critical limitation: fully self-consistent QPGW, in principle, overestimates band gaps due to its lack of vertex corrections, HSE06 orbitals should be used cautiously as input for G$_0$W$_0$ calculations in such highly correlated systems because such overestimation can lead to a misinterpretation of the material's electronic properties. Additionally, our analysis revealed substantial differences in charge densities between PBE and HSE06 across all chromium-based 2D materials studied, including simpler systems such as CrS$_2$, CrHS$_2$, and CrO$_2$. Among these, the overestimation of the band gap by G$_0$W$_0$@HSE06 was observed only in magnetic CrHS$_2$. In contrast, nonmagnetic CrS$_2$ and CrO$_2$ exhibited behavior akin to sp-bonded materials, underscoring the conjoint impact of magnetism and correlation effects in these systems.

Interestingly, G$_0$W$_0$@PBE demonstrates similar behavior across all tested materials when contrasted with fully Green's function self-consistent QPGW$_0$@PBE, where the screening is not updated.
Our results allow us to conclude that with careful treatment, G$_0$W$_0$@PBE remains a viable approach for achieving accurate and reliable results, even for complex materials such as chromium-based systems.

\begin{suppinfo}
Cr$_2$CF$_2$ conformers, phonon band structure, input DFT calculations, charge densities differences of CrS$_2$, CrHS$_2$, and CrO$_2$, integration of charge densities differences, the convergence of G$_0$W$_0$ calculations, and iterations evolution in full self-consistent quasiparticle QPGW calculations. 
\end{suppinfo}

%%%%%%%%%%%%%%%%%%%%%%%%
%%% ACKNOWLEDGEMENTS %%%
%%%%%%%%%%%%%%%%%%%%%%%%
\begin{acknowledgement}
This work was supported by the Czech Science Foundation (21-28709S) and the European Union under the LERCO project (number CZ.10.03.01/00/22\_003/0000003) via the Operational Programme Just Transition. 
The computations were performed at IT4Innovations National Supercomputing Center (e-INFRA CZ, ID:90140). We thank Matúš Dubecký for helpful discussions.
\end{acknowledgement}

%%%%%%%%%%%%%%%%%%%%
%%% BIBLIOGRAPHY %%%
%%%%%%%%%%%%%%%%%%%%
\bibliography{refs}

\end{document}